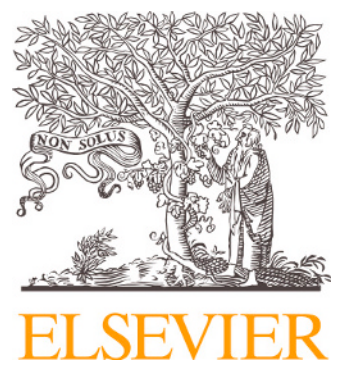

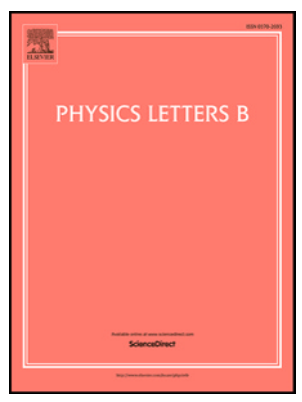

Letter

# Classical spacetime as a gravitational condensate: USMEG-EFT emergence in comparison to Verlinde's entropic gravity

Farrukh A. Chishtie [a,b,c]

[a] Peaceful Society, Science and Innovation Foundation, Vancouver, BC, Canada
[b] Department of Occupational Science & Occupational Therapy, Faculty of Medicine, University of British Columbia, Vancouver, BC, Canada
[c] Faculty of Land and Food Systems, University of British Columbia, Vancouver, BC, Canada

ABSTRACT

We present a derivation showing that classical spacetime geometry is a *gravitational condensate* within the Unified Standard Model with Emergent Gravity–Effective Field Theory (USMEG-EFT): the background metric $\bar{g}_{\mu\nu}$ is the vacuum expectation value of the quantum metric operator, existing as an ordered phase below the gravitational breakdown scale $\Lambda_{\rm grav} \sim 10^{18}$ GeV. The condensed phase is characterized by a nondegenerate metric expectation value, a diffeomorphism-invariant criterion whose controlled realization is confined to scales below $\Lambda_{\rm grav}$. Three convergent quantum-field-theoretic analyses within the framework, namely canonical covariance breakdown [1], one-loop renormalization group analysis [2], and conditional BRST closure [3], identify this scale as the boundary of the controlled geometric description. We show that the condensate order parameter, defined via the Legendre transform of the Lagrange multiplier path integral [4,5], satisfies a quantum-corrected saddle-point equation whose one-loop correction grows to the size of the tree term at $\Lambda_{\rm grav}$, beyond which the framework supplies no controlled nondegenerate solution; we interpret this boundary as the onset of condensate dissolution. Because the Lagrange multiplier constraint acts on virtual graviton fluctuations and terminates the gravitational sector exactly at one loop, this boundary arises from a finite, closed set of quantum corrections, in contrast to the gradual failure of the derivative expansion in the standard effective field theory treatment of general relativity [6]. This mechanism is fundamentally distinct from, and resolves known inconsistencies of, Verlinde's entropic gravity programme [7,8]: gravity is not an entropic statistical force; it is a gauge-mediated interaction whose *classical geometric description* condenses from quantum fluctuations below $\Lambda_{\rm grav}$. The graviton remains a well-defined spin-2 quantum field with exactly two transverse-traceless polarization modes, a parameter-free structural requirement consistent with LIGO-Virgo-KAGRA pure-polarization null tests [9,10].

## 1. Introduction

The hypothesis that spacetime geometry is emergent rather than fundamental has inspired two research programmes that share a surface vocabulary but are separated by every substantive physical and mathematical commitment.

In the entropic gravity programme [7,8], gravity is a statistical force arising from entropy gradients on holographic screens: $F = T\,\partial S/\partial x$. The entropy $S$ is assigned to arbitrary surfaces via the Bekenstein–Hawking area law; the graviton is not a primary degree of freedom; and the derivation proceeds at the level of thermodynamic heuristics rather than quantum field theory. The emergent quantity is gravity itself.

In the Unified Standard Model with Emergent Gravity–Effective Field Theory (USMEG-EFT) [1–3], the emergent quantity is categorically different: it is *classical geometry*. The graviton is a fully quantum, gauge-mediated spin-2 field; what emerges is the classical background metric $\bar{g}_{\mu\nu}$ as the expectation value of the quantum metric operator, a condensate that forms below $\Lambda_{\rm grav}$ and, we argue, dissolves above it, in the sense made precise in Section 3.4.

The seeds of this condensate picture were laid already in ref. [1]. That work established through canonical quantization that classical general covariance breaks down quantum mechanically above $\Lambda_{\rm grav}$, demonstrating that classical geometry has a *finite domain of validity* bounded above by a critical scale. This is precisely the defining hallmark of an ordered condensate phase: the background metric is not a universal description valid at all energies, but a phase-bounded structure whose existence is contingent on remaining below a critical threshold. Reference [2] extended this through renormalization group analysis, showing logarithmic running of quantum corrections that signals the EFT character of four-dimensional general relativity and places the one-loop breakdown at the same scale. Reference [3] supplied the path-integral confirmation through the demonstration that BRST closure in

*E-mail address:* fachisht@uwo.ca

gravity is conditional on the background, in contrast to its unconditional validity in Yang–Mills theory. Three convergent analyses, interrogating respectively the constraint algebra, the renormalization group, and the BRST structure of the same framework, identified the same boundary scale; what remained was to supply the condensate formalism that makes the mechanism explicit at the level of quantum field theory. That is the task of the present letter.

The effective field theory character of four-dimensional general relativity below the Planck scale is well established through the work of Donoghue and collaborators [6,11]. What USMEG-EFT adds to this picture is structural: the Lagrange multiplier constraint terminates the gravitational loop expansion exactly at one loop, so the breakdown at $\Lambda_{\rm grav}$ is governed by a finite and closed set of quantum corrections rather than by the leading edge of an infinite tower of counterterms. It is this sharpness, together with the order parameter constructed below, that justifies describing $\Lambda_{\rm grav}$ as the boundary of an ordered geometric phase.

The existence, location, and stability of the gravitational condensate follow from calculations within the Lagrange multiplier gravity framework of Brandt et al. [4], McKeon et al. [5]. A Legendre transform defines the order parameter through the expectation value of the metric operator; a quantum-corrected saddle-point equation governs its dynamics; and a stability analysis of the one-loop effective action locates the boundary of the controlled condensate regime at $\Lambda_{\rm grav}$ with no free parameters. The structural analogy with the electroweak Higgs condensate is close at the level of the quantum field theory formalism and equally illuminates what is genuinely novel: the Higgs condensate gives mass to fields that already propagate on a pre-existing spacetime, while the gravitational condensate *creates the spacetime stage itself*.

This letter presents the condensate mechanism in full, compares it with Verlinde's programme on every key dimension, situates the construction within the broader spacetime-as-condensate literature, and identifies the experimental anchor, a discrete graviton polarization count of exactly two, as a parameter-free structural requirement of the framework that current pure-polarization null tests support [9,10].

## 2. Untenability of the entropic gravity programme

Three independent results establish that Verlinde's programme is untenable as a fundamental theory. These are outlined as follows.

### 2.1. First-law violation for holographic screens

Wang and Braunstein [12] tested whether the surfaces central to Verlinde's construction obey an analogue of the first law of thermodynamics. The result is definitive: the first law holds on genuine horizons and remains a good approximation for stretched horizons immediately adjacent to them, but *fails for all ordinary surfaces* away from horizons except in the special case of spherical symmetry. Since Verlinde's derivation of $F = T\,\partial S/\partial x$ requires the first law to hold on generic holographic screens [7], this failure directly undermines the central thermodynamic assumption of the entire programme.

### 2.2. Causal inversion and circular reasoning

Gao [13] demonstrated that in the interaction between a holographic screen and a test particle, the entropy change of the screen is not the *cause* of the displacement but its *effect*: the gravitational interaction moves the particle, and the displacement then alters the screen entropy. Identifying the entropy gradient as the source of gravitational force therefore presupposes gravity, rendering the derivation circular. A further inconsistency noted in [13] is that the Unruh temperature $T = a\hbar/2\pi c$, itself a result of relativistic quantum field theory, is used to derive Newton's second law $F = ma$, which is non-relativistic: the relativistic framework that justifies the thermodynamic input is logically prior to, and incompatible with, the Newtonian output being derived.

### 2.3. Quantum decoherence and ultra-cold neutron states

Kobakhidze [14] showed that if gravity were entropic, gravitationally bound quantum states could not maintain coherence: entropic forces are inherently dissipative and cannot support the conservative potential wells that coherent quantum states require. The GRANIT experiment [15] directly refutes this prediction, observing discrete quantized energy levels $E_n \propto z_n^{2/3}$ of ultra-cold neutrons in the Earth's gravitational field, where $z_n$ are the zeros of the Airy function, with energy separations $\Delta E_n \sim 10^{-12}$ eV consistent with a conservative quantum potential to within experimental precision. Entropic gravity, taken literally, predicts these states should not exist. Attempts to rescue the programme via a decoherence parameter $\sigma$ [16] introduce a free parameter and cannot reproduce the observed Airy-function spectrum without fine-tuning that defeats the purpose of a derivation from first principles.

## 3. The gravitational condensate in USMEG-EFT

### 3.1. Order parameter and its definition

The fundamental dynamical variable in USMEG-EFT is the full metric $g_{\mu\nu}$, decomposed in the background field formalism [2,5] as

$$g_{\mu\nu} = \bar{g}_{\mu\nu} + \kappa\phi_{\mu\nu}, \tag{1}$$

where $\bar{g}_{\mu\nu}$ is a classical background and $\phi_{\mu\nu}$ represents quantum fluctuations with $\kappa^2 = 16\pi G_N$.

The condensed geometric phase is characterized by the expectation value of the quantum metric operator itself,

$$\langle \hat{g}_{\mu\nu}\rangle_{\mu<\Lambda_{\rm grav}} = \bar{g}_{\mu\nu}, \qquad \det \bar{g}_{\mu\nu} \neq 0, \tag{2}$$

that is, by the existence of a *nondegenerate* metric expectation value. Nondegeneracy of $\langle \hat{g}_{\mu\nu}\rangle$ is a diffeomorphism-invariant criterion: it makes no reference to a fiducial metric or to a coordinate choice, and it correctly classifies Minkowski space, for which $\bar{g}_{\mu\nu} = \eta_{\mu\nu}$ is nondegenerate, as a condensed classical spacetime. The disordered phase is defined by the loss of this property: above $\Lambda_{\rm grav}$ the framework supplies no controlled nondegenerate expectation value, while quantum correlations $\langle \phi_{\mu\nu}\phi_{\rho\sigma}\rangle \neq 0$ are hypothesized to persist (Section 5).

This characterization also identifies the symmetry-breaking pattern that a Landau-type description requires. In the pre-geometric phase the theory possesses full local frame and diffeomorphism invariance acting on a state with no distinguished geometric structure. A nondegenerate metric (equivalently, vierbein) expectation value breaks this invariance down to the isometry group of the background, in direct structural parallel with the electroweak condensate $\langle H\rangle$ breaking $SU(2)\times U(1) \to U(1)_{\rm em}$. Such symmetry breaking by a nondegenerate vierbein expectation value is a recognized mechanism in first-order formulations of gravity [17,18]. Two caveats attach to this parallel and we state them explicitly. First, local gauge symmetries do not break spontaneously in the strict sense, by Elitzur's theorem [19], and the phrase breaking of diffeomorphism invariance is heuristic here as it is in the electroweak case: the invariant content is the existence of a nondegenerate expectation value whose stabilizer is the isometry group of the background, a statement made precise in a gauge-fixed description. We emphasize that the mechanism at work in USMEG-EFT is in any case not spontaneous breaking, and so does not conflict with Elitzur's theorem: the symmetry is exact at the classical level, and it is the quantum corrections that fail to preserve it as the boundary is approached, as expressed by the conditional BRST closure of Eq. (18) and by the canonical covariance breakdown of Ref. [1]. The symmetry separating the phases is broken by quantum corrections, not spontaneously; the Landau-type language of Table 1 describes the phases themselves, while the mechanism that bounds the ordered phase is quantum mechanical. Second, the second-order formalism employed throughout this letter presupposes a nondegenerate background and therefore cannot represent the disordered phase as a configuration of the theory as constructed: the decomposition (1), the constraint (7), and the heat-kernel results below are all defined on nondegenerate $\bar{g}_{\mu\nu}$. A positive representation of the disordered phase requires first-order vierbein variables, in which degenerate configurations lie inside the field space; Ref. [17] supplies a symmetry-breaking mechanism in precisely those variables, while Ref. [18] demonstrates, in 2+1 dimensions, that dynamics at degenerate configurations can be sensible, an existence proof rather than a four-dimensional construction. Within the present programme the positive characterization of the disordered phase is the hypothesis supplied by PSEP (Section 5). The two-phase picture is therefore a framework-level hypothesis motivated by the boundary analysis of Section 3.4, not a result derived within the second-order formalism of this letter, which characterizes the ordered phase and locates the boundary of its controlled description. For computational convenience we retain the fluctuation variable

$$\Phi_{\mu\nu} \equiv \langle \hat{\phi}_{\mu\nu} \rangle, \tag{3}$$

which is nonvanishing when the fluctuation is quantized about a reference configuration $g^{\text{ref}}_{\mu\nu}$ distinct from the true saddle point, $\hat{g}_{\mu\nu} = g^{\text{ref}}_{\mu\nu} + \kappa\hat{\phi}_{\mu\nu}$, so that $\Phi_{\mu\nu} = \kappa^{-1}(\bar{g}_{\mu\nu} - g^{\text{ref}}_{\mu\nu})$ measures the displacement of the condensate from that reference. When the expansion point coincides with the saddle point, $\Phi_{\mu\nu} = 0$ by construction, and the invariant content of the phase resides entirely in the nondegeneracy of $\langle \hat{g}_{\mu\nu} \rangle$ in Eq. (2). We do not exhibit in this letter an effective potential with two competing minima; the stability analysis of Section 3.4 locates the boundary of the ordered phase, and the construction of the effective potential across the boundary is deferred to a comprehensive companion paper.

Following [5], the background is defined via the Legendre transform of the generating functional $W[j, J]$ of the full theory including the Lagrange multiplier field:

$$\bar{g}_{\mu\nu} = \left. \frac{\partial W[j, J]}{\partial j^{\mu\nu}} \right|_{j=J=0}. \tag{4}$$

The background is therefore not postulated; it is the saddle-point of the quantum effective action. The condensate interpretation is not a reinterpretation of known results but the natural field-theoretic reading of the Legendre structure, made possible precisely because [1] identified the finite domain below which this saddle-point is stable.

### 3.2. Path-integral derivation of the condensate equation

The Lagrange multiplier action is [4], written in the background field decomposition (1), with the multiplier coupled to the *linearized* Einstein operator evaluated on the background,

$$S_{\text{LM}} = S_{\text{EH}}[\bar{g} + \kappa\phi] + \frac{1}{\kappa^2} \int d^4x \sqrt{-\bar{g}}\, \lambda^{\mu\nu} \mathcal{G}^{(1)}_{\mu\nu}[\bar{g}; \phi], \tag{5}$$

where $\mathcal{G}^{(1)}_{\mu\nu}[\bar{g}; \phi]$ denotes the Einstein tensor expanded to first order in the quantum fluctuation $\phi_{\mu\nu}$ about the general background $\bar{g}_{\mu\nu}$.

The partition function is

$$Z[\bar{g}] = \int \mathcal{D}\phi_{\mu\nu}\, \mathcal{D}\lambda^{\mu\nu} \exp\left[ \frac{i}{\hbar} \left( S_{\text{EH}}[\bar{g} + \kappa\phi] + \frac{1}{\kappa^2} \int \sqrt{-\bar{g}}\, \lambda^{\mu\nu} \mathcal{G}^{(1)}_{\mu\nu}[\bar{g}; \phi] \right) \right]. \tag{6}$$

Integrating over $\lambda^{\mu\nu}$ first yields the functional delta function [4,20]:

$$\int \mathcal{D}\lambda^{\mu\nu}\, e^{i(\kappa^2\hbar)^{-1} \int \sqrt{-\bar{g}}\, \lambda^{\mu\nu} \mathcal{G}^{(1)}_{\mu\nu}[\bar{g};\phi]} = \delta\left[ \mathcal{G}^{(1)}_{\mu\nu}[\bar{g}; \phi] \right]. \tag{7}$$

A comment on the path-integral measure is required here, since it fixes the normalization of the one-loop results below. In the original Lagrange multiplier formulation, with the plain measure $\mathcal{D}\phi_{\mu\nu}\, \mathcal{D}\lambda^{\mu\nu}$, the integration over $\phi_{\mu\nu}$ against the functional delta function (7) produces an inverse functional Jacobian, $\det^{-1}$ of the quadratic fluctuation operator, so that the one-loop effective action is twice the standard Einstein gravity value [21]. Throughout this letter we instead adopt the refined measure of Refs. [5,22], which contains the compensating factor $\det^{1/2}\left[\delta^2 S_{\text{LM}}/\delta\phi\, \delta\phi\right]$, exponentiated through two fermionic and one bosonic ghost field. The delta function then contributes $\det^{-1}$, the measure factor contributes $\det^{1/2}$, and the net one-loop generating functional is the standard $\det^{-1/2}$. This choice restores unitarity of the constrained theory [22] and fixes the unmodified coefficients appearing in Eqs. (10) and (11); retaining the original measure would instead double the one-loop coefficients and shift the phase boundary of Section 3.4 by a factor $1/\sqrt{2}$. We emphasize the placement of the constraint, which is essential to the consistency of everything that follows: the functional delta function acts on the *virtual graviton fluctuations* $\phi_{\mu\nu}$ in the path-integral measure, not on the background metric and not on the full metric. The background $\bar{g}_{\mu\nu}$ remains a free functional argument of the effective action, ranging over general sub-Planckian curved configurations with expansion parameter $\kappa^2 \bar{R}$; it is not restricted to Ricci-flat solutions. In the full unification programme the background is in any case sourced by Standard Model matter through the coupled field equations [2,5], so that physically relevant backgrounds carry $\bar{R} \neq 0$. The curvature-squared structures appearing in the one-loop effective action below are therefore evaluated on configurations that the theory contains, and the stability analysis of Section 3.4 probes the response of the effective action as a functional of this general background.

Expanding $S_{\text{LM}}$ to quadratic order in $\phi_{\mu\nu}$ via (1):

$$\begin{aligned} S_{\text{LM}} = S_{\text{EH}}[\bar{g}] &+ \int d^4x \sqrt{-\bar{g}} \left. \frac{\delta S_{\text{EH}}}{\delta g^{\mu\nu}} \right|_{\bar{g}} \kappa\phi^{\mu\nu} \\ &+ \frac{\kappa^2}{2} \int d^4x\, \phi^{\mu\nu} \mathcal{O}_{\mu\nu\rho\sigma}[\bar{g}]\, \phi^{\rho\sigma} + \mathcal{O}(\kappa^3) \\ &+ \frac{1}{\kappa^2} \int d^4x \sqrt{-\bar{g}}\, \lambda^{\mu\nu} \mathcal{G}^{(1)}_{\mu\nu}[\bar{g}; \phi], \end{aligned} \tag{8}$$

where $\mathcal{O}_{\mu\nu\rho\sigma}[\bar{g}]$ is the quadratic graviton kinetic operator. The linear term in (8) vanishes in the effective action formalism [2], because $\bar{g}_{\mu\nu}$ is defined as the field configuration that extremizes $\Gamma[\bar{g}]$:

$$\frac{\delta\Gamma[\bar{g}]}{\delta\bar{g}^{\mu\nu}} = 0. \tag{9}$$

Eq. (9) is the **condensate equation**, the quantum-corrected Einstein equation that determines the expectation value $\bar{g}_{\mu\nu}$.

The constraint (7) confines the theory to one-loop order: multi-loop diagrams require off-shell graviton propagators with $\mathcal{G}^{(1)}_{\mu\nu} \neq 0$, which the functional delta function excludes [4]. This is the Lagrange multiplier mechanism's most consequential structural feature. It is also the technical realization of the condensate stability: the quantum corrections are bounded to one loop by the constraint, preventing the unbounded quantum noise that would, in its absence, destroy the ordered phase. The Lagrange multiplier field renormalization absorbs the resulting divergences without running $G_N$:

$$\lambda^{\mu\nu}_{\text{ren}} = \lambda^{\mu\nu} - \frac{\kappa^2}{(4\pi)^2} \left[ \frac{7}{20} \bar{G}^{\mu\nu} + \frac{1}{120} \bar{G}\, \bar{g}^{\mu\nu} \right], \tag{10}$$

where $\bar{G} = \bar{g}^{\mu\nu} G_{\mu\nu}[\bar{g}]$. The non-running of $G_N$ is therefore a structural consequence of condensate physics: Newton's constant is a property of the ordered condensed phase, not a running coupling. This contrasts sharply with standard perturbative quantum gravity and with every asymptotic safety programme, in which $G_N$ necessarily runs.

### 3.3. One-loop quantum correction to the condensate

The one-loop effective action, computed via heat kernel methods in dimensional regularization [2], is

$$\Gamma^{(1)}_{\text{fin}} = \frac{1}{(4\pi)^2} \ln\left( \frac{\mu}{\Lambda} \right) \int d^4x \sqrt{-\bar{g}} \left[ \frac{1}{120} \bar{R}^2 + \frac{7}{20} \bar{R}_{\mu\nu} \bar{R}^{\mu\nu} \right], \tag{11}$$

where the coefficients 1/120 and 7/20 are the classic 't Hooft–Veltman one-loop results [23], here arising within the constrained one-loop-exact structure. Here $\mu$ is the renormalization scale and $\Lambda$ the reference scale introduced by dimensional regularization at which the finite one-loop coefficients are defined [2]. The quantum-corrected condensate equation at one loop becomes

$$G_{\mu\nu}[\bar{g}] = -8\pi G_N \langle T^{(1)}_{\mu\nu}\rangle, \tag{12}$$

where the one-loop quantum stress tensor is

$$\begin{aligned}\langle T^{(1)}_{\mu\nu}\rangle &= -\frac{2}{\sqrt{-\bar{g}}}\frac{\delta\Gamma^{(1)}_{\text{fin}}}{\delta\bar{g}^{\mu\nu}} \\ &= -\frac{\ln(\mu/\Lambda)}{(4\pi)^2}\Bigg[\frac{1}{30}\bar{R}\bar{R}_{\mu\nu} - \frac{1}{120}\bar{g}_{\mu\nu}\bar{R}^2 + \frac{7}{5}\bar{R}_{\mu\alpha\nu\beta}\bar{R}^{\alpha\beta} \\ &\quad - \frac{7}{20}\bar{g}_{\mu\nu}\bar{R}_{\alpha\beta}\bar{R}^{\alpha\beta} - \frac{11}{15}\nabla_\mu\nabla_\nu\bar{R} + \frac{23}{60}\bar{g}_{\mu\nu}\Box\bar{R} \\ &\quad + \frac{7}{10}\Box\bar{R}_{\mu\nu}\Bigg].\end{aligned} \tag{13}$$

All coefficients in Eq. (13) have been verified symbolically by an independent Euler–Lagrange derivation of the variations of $\sqrt{-\bar{g}}\,\bar{R}^2$ and $\sqrt{-\bar{g}}\,\bar{R}_{\mu\nu}\bar{R}^{\mu\nu}$ on an anisotropic background, and the resulting tensor consistently vanishes on maximally symmetric configurations, as required by the on-shell triviality of curvature-squared variations there. The condensate expectation value receives the one-loop correction

$$\Phi_{\mu\nu} = \Phi^{(0)}_{\mu\nu} + \delta\Phi^{(1)}_{\mu\nu}, \quad \delta\Phi^{(1)}_{\mu\nu} = -\left[\frac{\delta^2 S_{\text{EH}}}{\delta\bar{g}^2}\right]^{-1}\frac{\delta\Gamma^{(1)}_{\text{fin}}}{\delta\bar{g}^{\mu\nu}}, \tag{14}$$

where $[\delta^2 S_{\text{EH}}/\delta\bar{g}^2]^{-1}$ is the graviton propagator evaluated on the background. For $\mu \ll \Lambda_{\text{grav}}$, Appelquist–Carazzone decoupling [24] ensures that the one-loop term, suppressed by $\bar{R}/M^2_{\text{Pl}}$ relative to the tree term, satisfies $\delta\Phi^{(1)}_{\mu\nu} \ll \Phi^{(0)}_{\mu\nu}$: the condensate is stable and classical geometry is a reliable description.

### 3.4. *Phase boundary at $\Lambda_{\text{grav}}$: loss of the controlled ordered phase*

The condensate is stable so long as the one-loop correction in (14) remains a perturbation to the tree-level term. The stability condition is

$$\left|\frac{\delta\Gamma^{(1)}_{\text{fin}}}{\delta\bar{g}^{\mu\nu}}\right| \ll \left|\frac{\delta S_{\text{EH}}}{\delta\bar{g}^{\mu\nu}}\right|. \tag{15}$$

Using $\delta S_{\text{EH}}/\delta\bar{g}^{\mu\nu} \sim M^2_{\text{Pl}}\bar{R}_{\mu\nu}/2$, with $M_{\text{Pl}}$ the reduced Planck mass, $M^2_{\text{Pl}} \equiv 2/\kappa^2 = (8\pi G_N)^{-1}$, and the leading terms from (13):

$$\frac{\delta\Gamma^{(1)}_{\text{fin}}}{\delta\bar{g}^{\mu\nu}} \sim \frac{\ln(\mu/\Lambda)}{(4\pi)^2}\cdot\mathcal{O}(\bar{R}^2)_{\mu\nu}. \tag{16}$$

We now probe the effective action, as a functional of the general background admitted by the constrained measure, on configurations with curvature scale $\bar{R} \sim \mu^2$; such backgrounds are legitimate arguments of $\Gamma[\bar{g}]$ since, as established after Eq. (7), the constraint restricts fluctuations rather than the background. The criterion (15) is a statement about the quantum corrections of the one-loop-exact gravitational sector relative to the tree term. It does not compare the magnitudes of classical Wilson coefficients in the effective action: curvature-squared terms induced by the matter sector, whatever their size, are couplings of the ordered phase that enter the condensate Eq. (12) without altering the quantum-hierarchy condition. The distinction is not optional. A matter-induced $R^2$ coefficient $c_2$ begins to dominate the Einstein–Hilbert term at the curvature threshold $\bar{R} = M^2_{\text{Pl}}/2c_2$, which for the value of $c_2$ required by the inflationary amplitude (Section 5) is $\bar{R} = 6M^2$, i.e. $\sqrt{\bar{R}} = \sqrt{6}\,M$, by construction of $c_2 = M^2_{\text{Pl}}/12M^2$. The observable Starobinsky epoch itself sits well above this threshold: Jordan-frame slow roll gives $\bar{R} \simeq 4NM^2 \approx 220\,M^2$ at $N = 55$, so that classical geometry manifestly persists in a regime where the $R^2$ term dominates the tree term by a factor $2N/3 \approx 37$. Dominance of a classical curvature-squared coupling therefore does not signal loss of a nondegenerate metric expectation value, and the phase boundary below is determined by the graviton sector alone. This assignment is moreover immaterial in practice: even if the derived matter-loop contribution, $c_2 \sim N_{\text{dof}}/(4\pi)^2 \approx 0.6$, is nonetheless retained in the criterion, solving the correspondingly modified condition self-consistently shifts the boundary only to $(1.8\text{–}2.1)\times 10^{18}$ GeV, of the same order as the range quoted below, so the conclusion is insensitive to where the matter-loop term is placed. The stability condition fails when

$$\frac{\mu^2\ln(\mu/\Lambda)}{(4\pi)^2} \sim \frac{M^2_{\text{Pl}}}{2} \implies \mu \sim \Lambda_{\text{grav}} \sim \frac{2\sqrt{2}\,\pi\,M_{\text{Pl}}}{\sqrt{\ln(\Lambda_{\text{grav}}/\Lambda)}}. \tag{17}$$

Solving self-consistently with the reduced Planck mass $M_{\text{Pl}} = 2.4\times 10^{18}$ GeV gives, up to order-unity and logarithmic factors, $\Lambda_{\text{grav}} \sim 10^{18}$ to $10^{19}$ GeV; for reference scales $\Lambda$ ranging from the TeV scale to $10^{16}$ GeV one finds $\Lambda_{\text{grav}} \approx 3.6\times 10^{18}$ to $8.3\times 10^{18}$ GeV, in agreement with all three analyses of Refs. [1–3]. At $\mu = \Lambda_{\text{grav}}$, the one-loop corrections become comparable to the tree-level Einstein–Hilbert term: the hierarchy on which the controlled description of the ordered phase rests fails, and, as discussed below, the framework supplies no gauge-unambiguous nondegenerate solution of the condensate Eq. (12) beyond this scale. We interpret $\Lambda_{\text{grav}}$ as the boundary of the condensed geometric phase.

We are careful about the status of this statement relative to the standard effective field theory picture [6], and about what the one-loop exactness of the gravitational sector does and does not license. In the unconstrained EFT, comparability of curvature-squared corrections with the Einstein–Hilbert term signals loss of predictivity because unknown higher orders then dominate. In USMEG-EFT there are no higher orders, and one might therefore suppose that the condensate Eq. (12), being closed, could simply be solved at curvature of order $\Lambda^2_{\text{grav}}$, with large but fully known corrections. The obstruction is that one-loop exactness is exactness of the loop series, not scheme independence of the extrapolated equation. The effective action entering Eq. (12) is an off-shell quantity whose gauge dependence is quantified by Eq. (19) below: the leading curvature-squared coefficient shifts by an order-unity amount between gauges, a consequence of the conditional BRST closure of Eq. (18), which fails precisely for background curvature that is not small in Planck units. Below $\Lambda_{\text{grav}}$ these gauge-dependent terms are parametrically suppressed relative to the tree term and the nondegenerate saddle point is a controlled, gauge-unambiguous prediction. At $\Lambda_{\text{grav}}$ the hierarchy fails and the gauge ambiguity of the saddle-point condition becomes of order unity simultaneously, so that solving the equation beyond the boundary yields backgrounds without invariant meaning within the present formulation. The claim of this letter is therefore that $\Lambda_{\text{grav}}$ is the boundary of the controlled ordered phase, arising from a finite, closed set of quantum corrections rather than the leading edge of an infinite tower. Whether this loss of control is accompanied by genuine loss of nondegenerate solutions, which would establish dissolution in the strict sense, and whether the boundary carries the thermodynamic hallmarks of a Landau transition, such as a non-analytic free energy or a diverging correlation length, are questions requiring the construction of the effective potential across the boundary, which is deferred to the comprehensive companion paper.

The BRST analysis supplies an independent, gauge-theoretic consistency check. In a fundamental gauge theory such as Yang–Mills, BRST nilpotency $s^2 = 0$ holds unconditionally, because the structure constants are field-independent, protecting physical observables from gauge dependence. In gravity, closure is conditional: the structure functions are field-dependent, and

$$s^2\phi_{\mu\nu} = \kappa^2\big[\bar{R}_{\mu\rho\nu\sigma} + \mathcal{O}(\kappa\phi\cdot\partial^2 c)\big]c^\rho c^\sigma + \cdots, \tag{18}$$

vanishing only for $\bar{R} = 0$ and $\kappa\phi \ll 1$ ("conditional BRST closure" [3]). Explicit computation of the one-loop effective action in two independent gauges [3] gives the gauge variation of the leading beta coefficient:

$$\Delta\beta_1 \equiv \beta_1^{\text{Gold}} - \beta_1^{\text{tHV}} = -\frac{119}{120} - \frac{1}{120} = -1. \tag{19}$$

**Table 1**
Gravitational condensate versus Higgs condensate.

| Feature | Higgs condensate | Gravitational condensate |
|---|---|---|
| Order parameter | $\langle H\rangle = v/\sqrt{2}$ | Nondegenerate $\langle \hat{g}_{\mu\nu}\rangle$; Eq. (2) |
| Broken symmetry | $SU(2)\times U(1)\to U(1)_{\rm em}$ | Local frame/diffeomorphism invariance → background isometries (heuristic; broken by quantum corrections, not spontaneously; see Section 3.1) [1,19] |
| Condensate eq. | $\partial V/\partial\phi = 0$ | $\delta\Gamma/\delta\bar{g}^{\mu\nu} = 0$ |
| Stability | $\mu^2 < 0$ in $V(\phi)$ | LM constraint; one-loop Eq. (15) |
| Phase boundary | $T_{\rm EW}\sim 160\,{\rm GeV}$ | $\Lambda_{\rm grav}\sim 10^{18}$ GeV; Eq. (17) |
| Disordered phase | $SU(2)\times U(1)$ restored | No classical geometry; pre-geometric (hypothesized; Section 5) |
| What dissolves | Particle masses | Classical spacetime |

We state the interpretive weight of this result with care, since off-shell effective actions are generically gauge-dependent and only on-shell or S-matrix quantities are guaranteed gauge-invariant. The relevant contrast is with Yang–Mills theory, where the leading beta coefficient, although extracted from an off-shell calculation, is independent of the gauge parameter, protected by unconditional BRST closure. In the gravitational case the leading logarithmic coefficient itself shifts by an order-unity amount between gauges, Eq. (19), which is consistent with conditional closure on a curved background. We present this not as an independent proof of the phase boundary but as a corroborating diagnostic: the primary weight of the argument is carried by the stability analysis of Eqs. (15)–(17), with the gauge-variation result and the canonical analysis of Ref. [1] serving as consistency checks that identify the same scale, and, as discussed above, with the gauge-variation result additionally delimiting the regime in which the saddle point of the condensate equation is gauge-unambiguous.

### *3.5. Structural analogy with the Higgs condensate*

The parallel with the electroweak Higgs condensate [25] is close at the level of the QFT formalism and illuminates what is genuinely novel. Table 1 summarizes the correspondence feature by feature, including the symmetry-breaking pattern identified in Section 3.1.

The key conceptual distinction is fundamental: the Higgs condensate gives mass to fields propagating on a pre-existing spacetime; the gravitational condensate *creates the spacetime stage on which all physics takes place*. No existing framework, including string theory, loop quantum gravity, asymptotic safety, or entropic gravity, has derived this creation mechanism from the quantum effective action of four-dimensional general relativity with a one-loop-exact gravitational sector, as is done in this work.

### *3.6. Relation to other condensate approaches to spacetime*

The idea that spacetime is a condensate has a developed literature, and it is important to situate the present construction within it. Group field theory condensate cosmology [26,27] builds macroscopic geometry from condensates of non-geometric quanta of a group field, with the emergence of a smooth metric and its dynamics to be demonstrated from the microscopic theory. The Dvali–Gomez programme [28] models black holes and, by extension, semiclassical spacetimes as bound states of soft gravitons at a quantum critical point, working at the level of a mean-field particle picture. Hu's condensate viewpoint and the associated stochastic gravity programme [29] treat classical spacetime as a collective, hydrodynamic-like limit of unspecified microscopic degrees of freedom. USMEG-EFT differs from all three in its starting point and in what is derived rather than modeled: the construction operates directly on the quantum effective action of four-dimensional general relativity, the condensed phase is identified with the saddle-point of the Legendre transform of the constrained path integral, and the location of the phase boundary follows from the one-loop-exact structure with no auxiliary microscopic postulates. The complementary relationship is natural: the approaches above propose candidate microscopic substrates, while USMEG-EFT fixes, from the infrared side, the structural properties that any such substrate must reproduce below $\Lambda_{\rm grav}$, and PSEP (Section 5) supplies the candidate ultraviolet completion within the present programme.

## 4. The graviton as organized gauge mediator

Within the condensed phase, $\phi_{\mu\nu}$ is a well-defined spin-2 quantum field propagating on $\bar{g}_{\mu\nu}$. Its renormalization-corrected propagator, derived from (11) [3], is

$$\Delta_{\mu\nu,\rho\sigma}(k) = \frac{i}{k^2+i\varepsilon}\Big[P^{(2)}_{\mu\nu,\rho\sigma} - \tfrac{1}{2}P^{(0)}_{\mu\nu,\rho\sigma}\Big]\left(1 + \frac{\kappa^2 k^2}{(4\pi)^2}\ln\frac{k^2}{\Lambda^2_{\rm grav}}\right), \tag{20}$$

where the spin-2 and spin-0 projection operators are

$$P^{(2)}_{\mu\nu,\rho\sigma} = \tfrac{1}{2}(\theta_{\mu\rho}\theta_{\nu\sigma} + \theta_{\mu\sigma}\theta_{\nu\rho}) - \tfrac{1}{3}\theta_{\mu\nu}\theta_{\rho\sigma}, \tag{21}$$

$$P^{(0)}_{\mu\nu,\rho\sigma} = \tfrac{1}{3}\theta_{\mu\nu}\theta_{\rho\sigma}, \tag{22}$$

with $\theta_{\mu\nu} = \eta_{\mu\nu} - k_\mu k_\nu/k^2$; the momentum-space representation is written on a locally flat patch of the general background, in which the projector decomposition is standard. The Lagrange multiplier constraint (7) enforces the transverse-traceless condition on-shell: only spin-2 modes propagate, and no scalar graviton is generated at any loop order within the condensed phase. The discrete, parameter-free structural requirement is therefore

$$N^{\rm grav}_{\rm dof} = \boxed{2} \qquad (\text{transverse-traceless: } h_+,\ h_\times). \tag{23}$$

This integer is not tuned; it is fixed structurally by the condensate.

We state the evidential status of this result conservatively. Two transverse-traceless polarizations is the prediction of classical general relativity, and consistency of gravitational-wave data with pure tensor polarizations therefore supports general relativity together with every framework that reduces to it at low energies. The polarization count is accordingly a parameter-free *consistency requirement* that USMEG-EFT satisfies structurally, through the constraint (7), and a *null test* against alternatives whose additional propagating modes would contribute vector or scalar polarizations: Einstein–Cartan theory in the class with independent, propagating torsional degrees of freedom (up to 6 modes; see the detailed comparative assessment in [30]), $f(R)$ gravity (3 modes), scalar-tensor theories (3 modes), and massive gravity (5 modes). The current observational situation is as follows. The GW170817 analysis yields Bayes factors exceeding $10^{20}$ to 1 in favor of pure tensor over pure vector or pure scalar polarizations [9], improved to approximately $10^{26}$ to 1 (log Bayes factor near 60) when the gamma-ray-burst jet-orientation prior is included [10]. These are pure-polarization null tests; discrimination of generic mixed-polarization content with a two-to-three-detector network remains limited, and the collaboration is appropriately careful on this point in its catalog-wide tests of general relativity [31,32]. Within these stated limits, the data are fully consistent with $N^{\rm grav}_{\rm dof} = 2$ and disfavor frameworks whose extra modes would produce detectable pure non-tensor content. We note the asymmetry of this test across the theory space: for frameworks in which the polarization count is a free or model-dependent quantity, the null tests constrain parameters, whereas for USMEG-EFT the count is a rigid structural output of the constraint (7) with no parameter available to adjust. The framework is therefore maximally exposed to this test: a single confirmed non-tensor detection would falsify it outright, and the current null results are precisely what its structure requires.

**Table 2**
Structural comparison between Verlinde's entropic gravity and USMEG-EFT emergent geometry.

| Feature | Verlinde entropic gravity | USMEG-EFT (this work) |
|---|---|---|
| Mechanism | Entropy gradient on holographic screen [7] | Condensation of metric expectation value below $\Lambda_{\rm grav}$; Eq. (12) |
| Order parameter | None defined | Nondegenerate $\langle \hat{g}_{\mu\nu} \rangle$; Eq. (2) |
| Thermodynamic consistency | First law fails for holographic screens [12] | No thermodynamic postulate invoked |
| Logical consistency | Circular: gravity presupposed via Unruh $T$ [13] | No circularity; all inputs from LM QFT |
| Graviton status | Not a primary degree of freedom | Spin-2 field $\phi_{\mu\nu}$; propagator (20) |
| Quantum coherence | Entropic noise destroys bound states [14] | Unitary propagator; GRANIT states preserved |
| Polarization count | Not derivable | $N_{\rm dof}^{\rm grav} = 2$; consistent with LVK pure-polarization null tests [9,10] |
| $G_N$ behaviour | Derived from screen equipartition; status unclear | Condensate property; does not run; Eq. (10) |
| Breakdown scale | Not derived | $\Lambda_{\rm grav} \sim 2\sqrt{2}\,\pi M_{\rm Pl}/\sqrt{\ln(\Lambda_{\rm grav}/\Lambda)}$; Eq. (17) |
| Pre-geometric phase | Not defined | PSEP: $\langle \phi_{\mu\nu} \rangle = 0$; spatial energy epoch [35] |
| Inflation | None | Starobinsky form of the action from condensate structure; amplitude normalization conjectural (Section 5); Eq. (24) |
| Experimental anchor | Galaxy lensing (disputed [37]) | LVK pure-polarization null tests; parameter-free [9,10] |
| Spacetime setting | AdS-inspired; de Sitter horizon | 4D asymptotically flat; physically realized |

It is worth stating precisely which aspects of the graviton sector the polarization data can and cannot probe. The observed strain is the coherent, classical limit of the quantum field $\phi_{\mu\nu}$, so polarization measurements test the mode content common to classical general relativity and to any consistent quantization of a massless spin-2 field, for which two helicity states follow from Poincaré invariance and gauge symmetry alone [33]. The polarization count therefore cannot separate USMEG-EFT from other frameworks with the correct massless spin-2 limit; what does separate them is the structure of quantum corrections. In the standard effective field theory treatment [6,11], the leading corrections are computable but sit at the edge of an infinite tower of counterterms, beginning with the Goroff–Sagnotti divergence at two loops [34], whose coefficients are free parameters fixed only by an unknown ultraviolet completion. In USMEG-EFT the constraint (7) never generates this tower: the two-loop amplitude is structurally absent, and the graviton sector is exhausted by the closed one-loop expressions (13) and (20). The renormalization-corrected propagator (20) is accordingly a *complete* quantum prediction rather than the leading term of an open series, and it is at this level, in the logarithmic momentum dependence of graviton propagation rather than in the mode count, that the framework is in principle distinguishable from the unconstrained effective field theory.

Verlinde's programme makes no prediction for $N_{\rm dof}^{\rm grav}$. Since the graviton is not a primary degree of freedom in entropic gravity, no propagator analogous to (20) is derivable from that framework, and it is silent on why gravitational-wave observatories observe exactly two polarization modes. This is not a technical omission that future development might remedy; it is a consequence of the framework's foundational architecture, in which no graviton field exists to be counted.

## 5. The pre-geometric phase and PSEP

Above $\Lambda_{\rm grav}$, the framework supplies no controlled nondegenerate metric expectation value and classical geometry is no longer a controlled description. The hypothesis developed in this section, which the considerations of Section 3.4 motivate but do not by themselves demonstrate, is that this boundary, at which quantum corrections cease to respect the classical symmetry, marks a genuine change of phase: that the physical substrate from which geometry condenses is *not geometric*. The Principle of Spatial Energy Potentiality (PSEP) [35] identifies this substrate as pre-geometric spatial energy: a quantum state of gravitational fluctuations $\phi_{\mu\nu}$ with $\langle \phi_{\mu\nu} \rangle = 0$ but $\langle \phi_{\mu\nu}\phi_{\rho\sigma} \rangle \neq 0$. Correlations exist; geometry does not.

In this scenario, the PSEP transition at $T \sim \Lambda_{\rm grav}$ is the cosmological phase transition at which the metric acquires a stable nondegenerate expectation value. The emergent one-loop action (11) contains the term $\frac{1}{120(4\pi)^2}\ln(\mu/\Lambda)\int\sqrt{-\bar{g}}\,\bar{R}^2$, which below $\Lambda_{\rm grav}$ combines with the Einstein–Hilbert term to produce an action of the Starobinsky form:

$$S_{\rm Star} = \frac{M_{\rm Pl}^2}{2}\int d^4x\sqrt{-g}\left(R + \frac{R^2}{6M^2}\right). \tag{24}$$

We are explicit about what is and is not derived here. The one-loop structure fixes the *functional form* of the action, and hence the shape of the inflationary predictions $n_s = 1 - 2/N$ and $r = 12/N^2$ as functions of the number of e-foldings, consistent with Planck 2018 constraints [36] for $N \approx 60$. It does not, from the graviton loop alone, reproduce the observed scalar amplitude: matching $A_s$ requires $M \approx 1.3 \times 10^{-5}\, M_{\rm Pl}$, corresponding to an $R^2$ coefficient of order $10^8$ to $10^9$, whereas the pure graviton contribution $\ln(\mu/\Lambda)/[120(4\pi)^2]$ is many orders of magnitude smaller. The origin of the required coefficient within USMEG-EFT is an open question, and we state its status without qualification: it is conjectural. One-loop Standard Model contributions to the $R^2$ coefficient are generically of order $N_{\rm dof}/(4\pi)^2 = \mathcal{O}(1)$, so that reaching $c_2 \sim 5 \times 10^8$ through the matter sector would require nonminimal couplings of order $\xi^2 \sim 10^9$, which the framework does not supply and which we do not introduce. Whether the unified framework generates the observed amplitude through a mechanism beyond minimal one-loop matter contributions is the subject of a dedicated companion computation in preparation, and until that computation is complete the amplitude matching should be regarded as an assumption external to this letter. The structural statement made here is therefore limited and precise: the Starobinsky *form* of the inflationary action is a consequence of the condensate structure of USMEG-EFT rather than an independent assumption, while the amplitude normalization is not derived in this work. We note that a large $c_2$, if realized, would not displace the phase boundary of Section 3.4, for the reason given there: the threshold $\sqrt{6}\,M$ at which such a coefficient begins to dominate the Einstein–Hilbert term lies below the curvature attained during the inflationary epoch itself, a fully geometric regime, and classical Wilson coefficients do not enter the quantum-hierarchy criterion. Verlinde's programme has no analogue of this: having placed gravity outside quantum field theory, it generates no inflationary dynamics and no connection to the CMB power spectrum.

## 6. Systematic comparison

Table 2 presents a systematic feature-by-feature comparison of Verlinde's entropic gravity programme and USMEG-EFT across every dimension on which the two frameworks differ.

## 7. Discussion and conclusion

The condensate picture of classical spacetime presented here has its roots in Ref. [1], where canonical quantization first demonstrated that general covariance is not universal but phase-bounded: classical geometry exists and is reliable only below $\Lambda_{\text{grav}}$, above which covariance breaks down quantum mechanically. That result established the necessary precondition for a condensate, namely a bounded ordered phase, without yet supplying the condensate formalism. The present letter completes the argument. The order parameter is the nondegenerate expectation value of the quantum metric operator, Eq. (2), a diffeomorphism-invariant characterization whose associated symmetry-breaking pattern parallels the electroweak case at the level of the phases, though not of the mechanism, which is quantum mechanical rather than spontaneous (Section 3.1); the condensate Eq. (12) is its quantum-corrected equation of motion; the stability condition (17) locates the boundary of the controlled ordered phase at the same $\Lambda_{\text{grav}} \sim 10^{18}$ GeV identified by three convergent analyses [1–3]; and the LM constraint (7), acting on the virtual graviton fluctuations in the path-integral measure over a general sub-Planckian background, stabilizes the condensate to one-loop order while non-perturbatively excluding all multi-loop corrections that would destroy the ordered phase. The sharpness of the boundary relative to the standard EFT breakdown [6] rests on this one-loop exactness: what fails at $\Lambda_{\text{grav}}$ is the hierarchy between the tree term and a finite, closed set of corrections, not the leading edge of an infinite tower, and the loss of hierarchy coincides with the loss of gauge control of the saddle point established in Section 3.4. Condensate dissolution in the strict sense remains the interpretation this structure motivates, with its demonstration deferred to the companion paper. All quantitative statements in this letter, including the complete one-loop stress tensor Eq. (13), the coefficient in Eq. (17), and the projector algebra underlying Eq. (20), have been verified with an independent symbolic computation suite.

The mechanism is incompatible with Verlinde's entropic programme at every level. The graviton is not absent, statistical, or dissipative: it is a well-defined spin-2 quantum field with propagator (20), carrying exactly two transverse-traceless modes. $G_N$ does not run: it is a property of the condensed phase, not a coupling constant. The first law is not violated because no thermodynamic postulate is invoked. And the framework delivers a specific, parameter-free structural requirement, namely exactly two polarization modes, that current pure-polarization null tests support [9,10] and that the entropic programme cannot derive and has not attempted to derive.

The Higgs analogy is instructive. Just as the electroweak condensate $\langle H \rangle = v/\sqrt{2}$ generates particle masses and defines the low-energy gauge structure below $T_{\text{EW}}$, the gravitational condensate $\langle \hat{g}_{\mu\nu} \rangle$ generates classical spacetime geometry and defines the low-energy gravitational structure below $\Lambda_{\text{grav}}$. The difference is profound and not merely quantitative: the Higgs condensate operates *within* a pre-existing spacetime; the gravitational condensate *creates* it. Above $\Lambda_{\text{grav}}$, the PSEP pre-geometric phase prevails, which is hypothesized as a quantum state with nonzero graviton correlators but zero metric expectation value, from which the cosmos emerged in the phase transition at the beginning of the inflationary epoch.

The conclusion is unambiguous: entropy is a consequence of gravitational geometry, not its cause.

## Declaration of competing interest

The authors declare that they have no known competing financial interests or personal relationships that could have appeared to influence the work reported in this paper.

## Acknowledgments

F.A.C. thanks D.G.C. McKeon for useful discussions on the nature of quantum gravity and the Peaceful Society, Science and Innovation Foundation for support.

## Data availability

The article contains all the data used in the article